\documentclass[reprint,
superscriptaddress,
showpacs,
nofootinbib,
prl,
aps,
floatfix,
]{revtex4-1}

\usepackage{amsmath,amssymb,amsfonts,amsthm}
\usepackage[english]{babel}
\usepackage{url}
\usepackage{bm}
\usepackage[latin1]{inputenc}
\usepackage{graphicx,rotating}
\usepackage[colorlinks=true,linkcolor=black, citecolor=blue]{hyperref}

\usepackage{natbib}

\begin{document}
\title{First-forbidden transitions in the reactor anomaly}
\author{L. Hayen}
\email[Corresponding author: ]{leendert.hayen@kuleuven.be}
\affiliation{Instituut voor Kern- en Stralingsfysica, KU Leuven, Celestijnenlaan 200D, B-3001 Leuven, Belgium}

\author{J. Kostensalo}
\affiliation{Department of Physics, University of Jyv\"askyl\"a, P.O. Box 35, FI-40014 University of Jyv\"askyl\"a, Finland}

\author{N. Severijns}
\affiliation{Instituut voor Kern- en Stralingsfysica, KU Leuven, Celestijnenlaan 200D, B-3001 Leuven, Belgium}

\author{J. Suhonen}
\affiliation{Department of Physics, University of Jyv\"askyl\"a, P.O. Box 35, FI-40014 University of Jyv\"askyl\"a, Finland}

%
%
%
%

\date{\today}
\begin{abstract}
We study the dominant forbidden transitions in the antineutrino spectra of the fission actinides from 4 MeV onward using the nuclear shell model. Through explicit calculation of the shape factor, taking into account Coulomb corrections, we show the expected changes on cumulative electron and antineutrino spectra. Compared to the usual allowed approximation this results in a minor decrease of electron spectra from 4 MeV and onward, whereas an increase of several percent is observed in antineutrino spectra. We show that, despite their limited number, forbidden transitions dominate the spectral flux for most of the experimentally accessible range. Based on the shell model calculations we attempt a parametrization of forbidden transitions and propose a spectral correction for all forbidden transitions. We enforce correspondence with the ILL dataset using a summation+conversion approach. When compared against modern reactor neutrino experiments, the resultant spectral change is observed to be of comparable magnitude and shape as the reported spectral shoulder, drastically decreasing the statistical significance of the latter.
\end{abstract}

\maketitle

For the past years, neutrino physics has seen a flurry of interest in the so-called reactor anomaly \cite{Mention2011, Mueller2011, Huber2011b}, a 6\% deficit in the experimentally observed antineutrino count rate relative to theoretical predictions. Together with more long-standing anomalies (LSND \cite{Athanassopoulos1998, Conrad2013}, GALLEX \cite{Kaether2010}), much theoretical interest has gone towards the possibility of one or more eV-scale sterile neutrinos \cite{Abazajian2012, Gariazzo2015}. With the availability of precision antineutrino spectra, however, all modern reactor neutrino experiments have also observed a spectral disagreement with respect to theoretical predictions in the region between 4 and 6 MeV \cite{An2016, Abe2014, Seo2018}. Up to now, this so-called shoulder has remained unexplained, and several possibilities have been proposed for its solution \cite{Mention2017, Hayes2016, Hayes2015, Huber2016a, Buck2017}. The role of (first) forbidden transitions in both the anomaly and shoulder has so far received limited study, either in parametrized \cite{Hayes2014} or microscopic treatments \cite{Fang2015}. Based on the behaviour of pseudoscalar ($\Delta J^\pi = 0^-$) transitions the forbidden influence has been estimated as negligible \cite{Hayes2015}, despite its flux dominance in the region of interest \cite{Sonzogni2015}. Here we investigate the influence of first-forbidden $\beta$ decays by calculating the shape factor of the dominant transitions in the region of interest using the nuclear shell model, and show its far-reaching consequences. A more extensive discussion will be published elsewhere \cite{Hayen2018TBP}.

We use the formalism of Behrens and B\"uhring \cite{Behrens1982} to properly describe the spectral shape, taking into account finite-size and Coulomb corrections at all levels. We write the $\beta$ spectrum as
\begin{equation}
    \frac{dN}{dW} = pW(W-W_0)^2F(Z, W)C(Z, W)K(Z, W)
    \label{eq:spectrum_shape_general}
\end{equation}
where $W = E/m_ec^2 + 1$ is the total $\beta$ energy, $p=\sqrt{W^2-1}$ the momentum, $W_0$ the spectral endpoint and $Z$ the proton number of the daughter. Additionally, $F(Z, W)$ is the Fermi function, $C(Z, W)$ the shape factor and $K(Z, W)$ additional higher-order effects \cite{Hayen2018}. In previous analyses \cite{Huber2011b, Dwyer2015} forbidden transitions were approximated as allowed, either using $C=1$ or including a linear weak magnetism correction so that $\mathrm{d}C/\mathrm{d}W = 0.67\%$ MeV$^{-1}$. In the so-called Huber-Mueller (H-M) case, all forbidden transitions were assumed to have a unique shape \cite{Mueller2011}. For clarity, we will make a comparison against the allowed approximation, and comment on the H-M approximation.

We write the generalized unique forbidden shape factor of order $L$ as  \cite{Behrens1969}
\begin{equation}
    C_U = \sum_{k=1}^L\lambda_k \frac{p^{2(k-1)}q^{2(L-k)}}{(2k-1)![2(L-k)+1]!},
    \label{eq:C_unique_forbidden}
\end{equation}
and for illustrative purposes we write the first-forbidden pseudoscalar and pseudovector shape factors using their dominant parts as
\begin{align}
    C_{0^-} &= 1 + \frac{2R}{3W}b + \mathcal{O}(\alpha Z R, W_0R^2),
    \label{eq:C_pseudoscalar}\\
    C_{1^-} &= 1 + aW + \mu_1 \gamma \frac{b}{W} + cW^2, \label{eq:C_pseudovector}
\end{align}
where $R$ is the nuclear radius, $\alpha$ is the fine-structure constant, $\gamma = \sqrt{1-(\alpha Z)^2}$, $q = W_0-W$ is the (anti)neutrino momentum and $\lambda_k$ a Coulomb correction function. Here $a, b$ and $c$ are (complex) combinations of nuclear matrix elements, corresponding to powers of $W$ $+1, -1$ and $+2$. Note that we have not used the simplified expressions of Eqs. (\ref{eq:C_pseudoscalar}) \& (\ref{eq:C_pseudovector}), but rather used the complete formulation as can, e.g., be found in Ref. \cite{Behrens1971}. As such, we additionally take into account finite-size effects and Coulomb corrections to the nuclear matrix elements. As cancellations can occur in the main matrix elements, the latter can eventually dominate the shape factor. The importance of these corrections can not be understated, in particular for the high masses of the actinide fission fragments \cite{Hayen2018TBP}. 

Based on the compilation in Ref. \cite{Sonzogni2015} we have selected 29 forbidden transitions, the properties of which are listed in Table \ref{tab:summary_transitions_4MeV_235U}. All transitions have a $\beta$ spectrum endpoint above 4 MeV meaning all contribute to the observed spectral shoulder. Using the experimental results obtained at the ILL \cite{Schreckenbach1981, Feilitzsch1982, Schreckenbach1985, Hahn1989, Haag2014}, the selected transitions correspond to at least 50\% of the cumulative electron flux in the region of interest (2-8 MeV), and exceed 65\% at 6 MeV for all fission actinides \cite{Hayen2018TBP}. The shell model calculations were performed using the shell model code NUSHELLX@MSU \cite{Brown2014}. For nuclei with $A<100$ the effective interaction glepn \cite{Mach1990} was adopted in a full model space consisting of the proton orbitals $0f_{5/2}-1p-0g_{9/2}$ and the neutron orbitals $1d-2s$. The $^{86}$Br and $^{89}$Br cases were calculated using the interaction jj45pna \cite{Machleidt2001, Lalkovski2013}, in the full model space spanned by the proton orbitals $0f_{5/2}-1p-0g_{9/2}$ and the neutron orbitals $0g_{7/2}-2s-1d-0h_{11/2}$. For the nuclei with $A=$133--142 the Hamiltonian jj56pnb \cite{Brown2012} was used in the full model space spanned by the proton orbitals $0g_{7/2}-1d-2s-0h_{11/2}$  and neutron orbitals $0h_{9/2}-1f-2p-0i_{13/2}$ for $A<139$, while for the heavier nuclei the proton orbital $0h_{11/2}$ and the neutron orbital $0i_{13/2}$ were kept empty due to the enormous dimensions of a full model space calculation. Uncertainties due to $g_A$ quenching and meson exchange currents (MEC) in pseudoscalar transitions in the relevant areas are discussed elsewhere \cite{Kostensalo2018, Kostensalo2017, Haaranen2017, Suhonen2017a, Hayen2018TBP, Bodenstein-Dresler2018}, and were accounted for in this study.

We have separately used the ENSDF \cite{ENSDF} and ENDF \cite{Chadwick2011} decay libraries. While the former likely suffers from multiple cases of the pandemonium effect \cite{Hardy1977}, the latter has been corrected to obtain improved agreement with experimental reactor data \cite{Gauld2014, Mendoza2017}. Unless mentioned explicitly, the results below are obtained using the ENDF/B-VIII.0 decay library with spin-parity information from ENSDF. Transitions with unknown spin-change are assumed allowed and unknown branching ratios are distributed equally from the remaining intensity \cite{Hayen2018TBP}.

\begin{table}[ht]
\caption{Summary of the calculated dominant forbidden transitions above 4 MeV. Here $Q_{\beta}$ is the ground-state to ground-state $Q$-value, $E_{ex}$ the excitation energy of the daughter level, BR the branching ratio of the transition normalized to one decay and FY the cumulative fission yield of $^{235}$U taken from the ENDF database \cite{Chadwick2011}.}
\begin{tabular}{c|cccccc}
\hline \hline
Nuclide & $Q_\beta$ & $E_{ex}$ & BR & $J^\pi_i \to J^\pi_f$ & FY & $\Delta J$ \\
& (MeV) & (MeV) & (\%) & & (\%) & \\
\hline
$^{89}$Br & 8.3 & 0 & 16 & $3/2^- \to 3/2^+$ & 1.1 & 0\\
$^{90}$Rb & 6.6 & 0 & 33 & $0^- \to 0^+$ & 4.5 & 0\\
$^{91}$Kr & 6.8 & 0.11 & 18 & $5/2^+ \to 5/2^-$ & 3.5 & 0\\
$^{92}$Rb & 8.1 & 0 & 95.2 & $0^- \to 0^+$ & 4.8 & 0\\
$^{93}$Rb & 7.5 & 0 & 35 & $5/2^- \to 5/2^+$ & 3.5 & 0\\
$^{94}$Y & 4.9 & 0.92 & 39.6 & $2^- \to 2^+$ & 6.5 & 0\\
$^{95}$Sr & 6.1 & 0 & 56 & $1/2^+ \to 1/2^-$ & 5.3 & 0\\
$^{96}$Y & 7.1 & 0 & 95.5 & $0^- \to 0^+$ & 6.0 & 0\\
$^{97}$Y & 6.8 & 0 & 40 & $1/2^- \to 1/2^+$ & 4.9 & 0\\
$^{98}$Y & 9.0 & 0 & 18 & $0^- \to 0^+$ & 1.9 & 0\\
$^{133}$Sn & 8.0 & 0 & 85 & $7/2^- \to 7/2^+$ & 0.1 & 0\\
$^{135}$Te & 5.9 & 0 & 62 & $(7/2-) \to 7/2^+$ & 3.3 & 0\\
$^{136m}$I & 7.5 & 1.89 & 71 & $(6^-) \to 6^+$ & 1.3 & 0\\
$^{136m}$I & 7.5 & 2.26 & 13.4 & $(6^-) \to 6^+$ & 1.3 & 0\\
$^{137}$I & 6.0 & 0 & 45.2 & $7/2^+ \to 7/2^-$ & 3.1 & 0\\
$^{138}$I & 8.0 & 0 & 26 & $0^+ \to 0^-$ & 1.5 & 0\\
$^{142}$Cs & 7.3 & 0 & 56 & $0^- \to 0^+$ & 2.7 & 0\\
$^{86}$Br & 7.3 & 0 & 15 & $(1^-) \to 0^+$ & 1.6 & 1 \\
$^{86}$Br & 7.3 & 1.6 & 13 & $(1^-) \to 2^+$ & 1.6 & 1\\
$^{87}$Se & 7.5 & 0 & 32 & $3/2^+ \to 5/2^-$ & 0.8 & 1\\
$^{89}$Br & 8.3 & 0.03 & 16 & $3/2^- \to 5/2^+$ & 1.1 & 1\\
$^{91}$Kr & 6.8 & 0 & 9 & $5/2^+ \to 3/2^-$ & 3.4 & 1\\
$^{134m}$Sb & 8.5 & 1.69 & 42 & $(7-) \to 6^+$ & 0.8 & 1\\
$^{134m}$Sb & 8.5 & 2.40 & 54 & $(7^-) \to (6^+)$ & 0.8 & 1\\
$^{140}$Cs & 6.2 & 0 & 36 & $1^- \to 0^+$ & 5.7 & 1 \\
$^{88}$Rb & 5.3 & 0 & 76.5 & $2^- \to 0^+$ & 3.6 & 2\\
$^{94}$Y & 4.9 & 0 & 41 & $2^- \to 0^+$ & 6.5 & 2\\
$^{95}$Rb & 9.2 & 0 & 0.1 & $5/2^- \to 1/2^+$ & 0.8 & 2\\
$^{139}$Xe & 5.1 & 0 & 15 & $3/2^- \to 7/2^+$ & 5.0 & 2\\
\hline \hline
\end{tabular}
\label{tab:summary_transitions_4MeV_235U}
\end{table}

\begin{figure}[ht]
    \centering
    \includegraphics[width=0.49\textwidth]{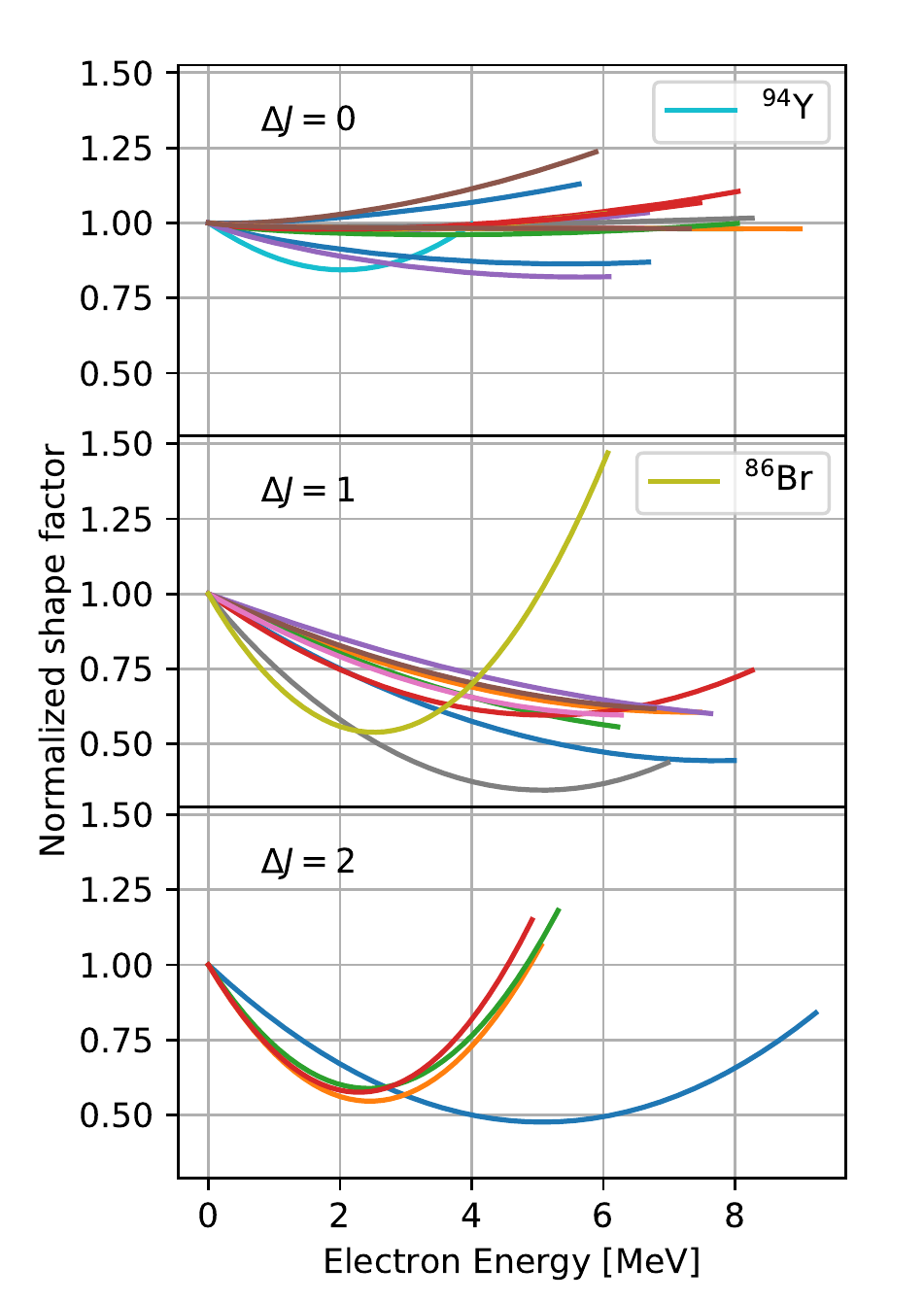}
    \caption{Overview of the calculated shape factors $C$ versus electron kinetic energy, categorized according the spin-parity change of the transition. For allowed transitions $C=1$. Each shape factor was normalized to its value at $E=0$. Results correspond to $g_A=0.9$ and $\epsilon_\text{MEC} = 1.4$, where applicable \cite{Kostensalo2018, Hayen2018TBP}. Two cases stand out: $^{94}$Y ($2^- \to 2^+$) and $^{86}$Br ($1^- \to 2^+$). Both contain strong admixtures of $\Delta J = 2$ operators, since both $2^+$ final states identify as vibrational excitations of the $0^+$ ground state.}
    \label{fig:shape_factors_collection}
\end{figure}

Figure \ref{fig:shape_factors_collection} shows the calculated shape factors categorized according to the change in spin-parity. It is immediately clear that all shape factors deviate significantly from unity and contain slopes much steeper than a weak magnetism term in the allowed approximation. The spin change is a good predictor for the calculated shape factor, with the exception of pseudoscalar, $\Delta J^\pi = 0^-$, transitions. From Eq. (\ref{eq:C_pseudoscalar}) its behaviour should be trivial with $|bR| \sim 10^{-2}$, yet large variations appear. As many of these transitions connect initial and final states with spins larger than zero, additional $\Delta J = 1, 2$ operators contribute non-negligibly. As such, in many cases the energy dependence is dominated by higher-order operators. Even though results appear to scatter around unity, the limited number of contributing branches forbids simple statistical averaging arguments.

\begin{figure}
    \centering
    \includegraphics[width=0.48\textwidth]{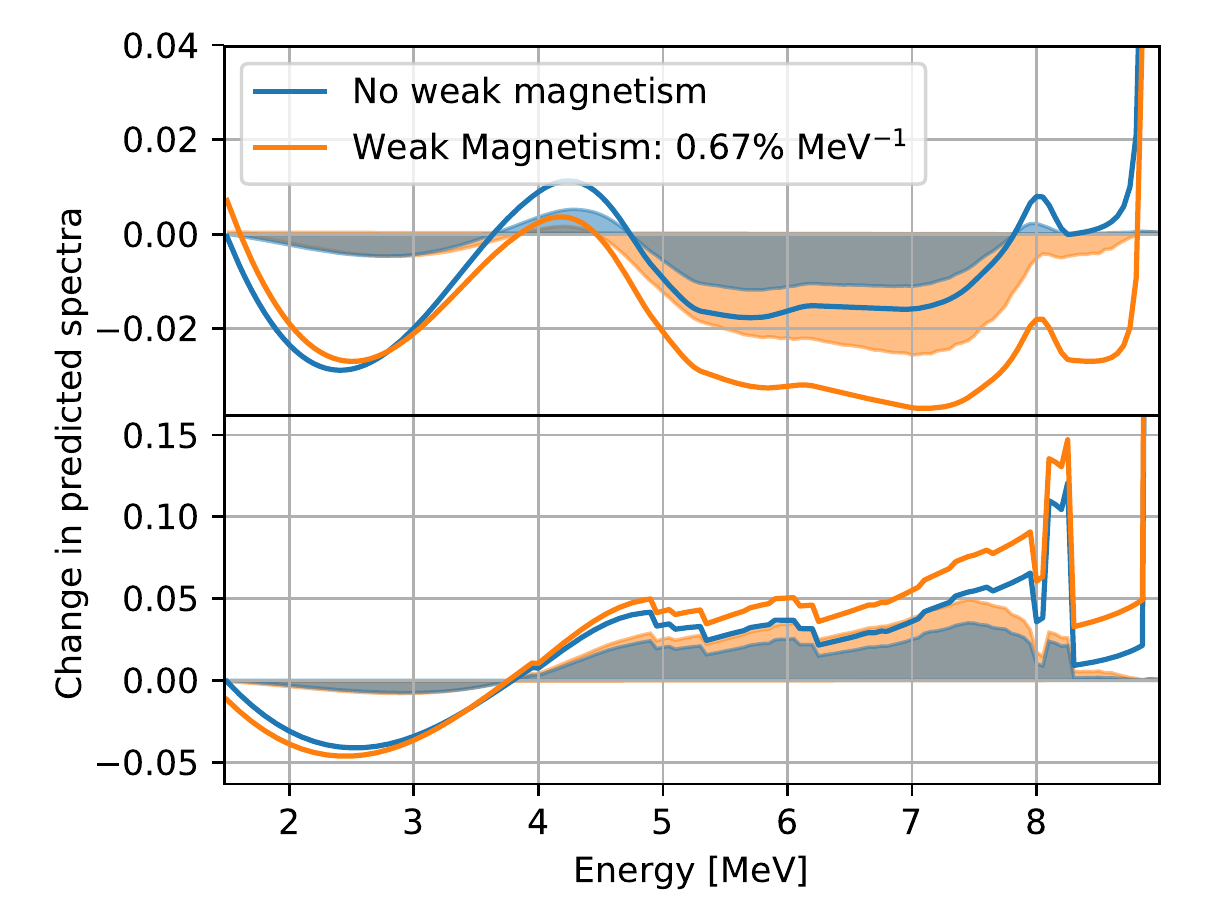}
    \caption{Top panel: Change in the predicted electron spectra of the considered transitions compared to the allowed approximation with an optional weak magnetism correction. Bottom panel: Change in the predicted antineutrino spectrum compared to the allowed approximation. Shaded areas correspond to the results multiplied by the total spectral contribution compared to experimental flux results. The energy axis refers to the kinetic energy of the electron (top) and antineutrino (bottom).}
    \label{fig:spectral_changes}
\end{figure}

Using the fission yields of the ENDF database \cite{Chadwick2011}, Fig. \ref{fig:spectral_changes} shows the change of both electron and antineutrino spectra compared to the allowed approximation with an optional weak magnetism correction. The shaded regions show the effective change to the total spectrum. Compared to the weak magnetism correction typically used \cite{Huber2011b, Mueller2011}, electron spectra see a modest 2\% decrease in the 4 to 8 MeV region. Cumulative antineutrino spectra, on the other hand, see a change of up to 5\% in the same region. The parabolic behaviour below 4 MeV is almost entirely attributable to first-forbidden unique decays (see, e.g., Fig. \ref{fig:shape_factors_collection}).

While a significant fraction of the spectral change occurs because of pseudovector ($\Delta J^\pi = 1^-$) transitions, inspection of Fig. \ref{fig:shape_factors_collection} should make it clear that even pseudoscalar transitions carry significant deviations from unity. Previous arguments for its neglect \cite{Hayes2015} have used $^{92}$Rb and $^{96}$Y as examples for their predictions, even though many important pseudoscalar transitions are not pure $0^- \to 0^+$ decays (see Fig. \ref{fig:shape_factors_collection}).

\begin{figure}[ht]
    \centering
    \includegraphics[width=0.48\textwidth]{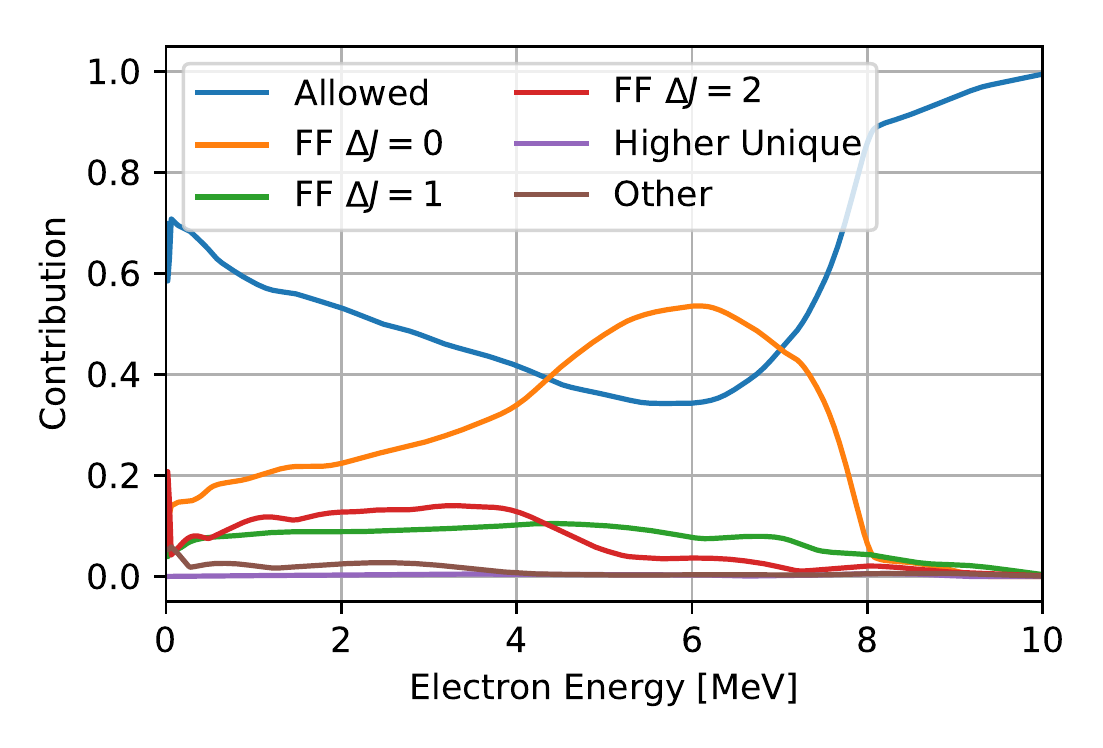}
    \caption{Constituents of the summed full $^{235}$U electron spectrum using the ENDF database \cite{Chadwick2011}. Here `FF' stands for first-forbidden and `Other' for non-unique transitions with $\Delta J \geq 2$. Behaviour past 10\,MeV is dominated by a low number of branches. Using decay information from ENSDF (not shown here) these features are strongly amplified, with the contribution of allowed decays reaching a minimum below 20\% around 5 MeV \cite{Hayen2018TBP}.}
    \label{fig:constituents_U235}
\end{figure}

While several compilations have been produced in the past \cite{Sonzogni2015, Hayes2018}, forbidden decays have typically been pushed to the background as they only make up about 30\% of the total number of transitions contributing to the total flux. Many of the large-endpoint transitions are, however, of forbidden nature due to the parity change of proton and neutron orbitals in the neutron-rich fission fragments. States of equal parity typically reside at excitation energies of several MeV with fragmented branching ratios, thereby pushing them out of the region of interest. In order to clarify these concerns, Fig. \ref{fig:constituents_U235} shows the constituents of the summed full $^{235}$U spectrum. It is immediately clear that allowed transitions, contrary to simple estimates, contribute less than 50\% in the entire experimentally interesting region. In the observed shoulder, in particular, forbidden transitions constitute more than 60\% of the total electron flux. The majority of these are pseudoscalar transitions, which in a pure $0^- \leftrightarrow 0^+$  transition show minimal deviation from an allowed equivalent. As shown above, however, this situation is not typical and subject to large higher-order contributions. Contributions from $\Delta J = 1, 2$ first-forbidden decays remain relatively constant throughout the entire spectrum up to 7 MeV, making up around 20\%. Given their strongly deviating shape factor as shown in Fig. \ref{fig:shape_factors_collection}, their influence cannot be understated.

In light of these results and the relatively uniform behaviour of the shape factors as calculated by the nuclear shell model, we attempt a simple parametrization. While the shape factor of pure pseudoscalar transitions (Eq. (\ref{eq:C_pseudoscalar})) is simple enough, the influence of higher-order operators prevents a physically insightful function description. As such, we simply fit the obtained shape factors with a general description as in Eq. (\ref{eq:C_pseudovector}) and similarly for pseudovector transitions. The shape factor of unique forbidden decays describes observed spectra within a few percent when properly taking into account Coulomb distortions. As such, we need no parametrization for $\Delta J \geq 2$ unique decays and instead simply use Eq. (\ref{eq:C_unique_forbidden}).

The parametrization then functions as follows \cite{Hayen2018TBP}. Each of the non-unique shape factors calculated by the nuclear shell model is fit using functions described above. For each spin-change ($\Delta J = 0, 1$), one obtains distributions of fit parameters. The resulting spread is dominated by differences between transitions rather than individual uncertainties arising from $g_A$ and $\epsilon_\text{MEC}$ ambiguity. Due to limited statistics, we use Gaussian kernel smoothing \cite{Scott1992} where we manually set the bandwidth to $h=2$. Our choice results in fit parameter distributions with conservative uncertainties where all shape factors of Fig. \ref{fig:shape_factors_collection} are contained within a $<2 \sigma$ window. Full spectra are then calculated in a Monte Carlo fashion, where for each non-unique first-forbidden transition fit parameters are obtained from the correlated parameter ensemble, with exception of the transitions numerically calculated in this work. Repeating this procedure many times results in a direct translation of the uncertainty of our parametrization into a spectral uncertainty. The numerous additional spectrum shape corrections in Eq. (\ref{eq:spectrum_shape_general}) are calculated using Ref. \cite{Hayen2018a}.

\begin{figure}[ht]
    \centering
    \includegraphics[width=0.48\textwidth]{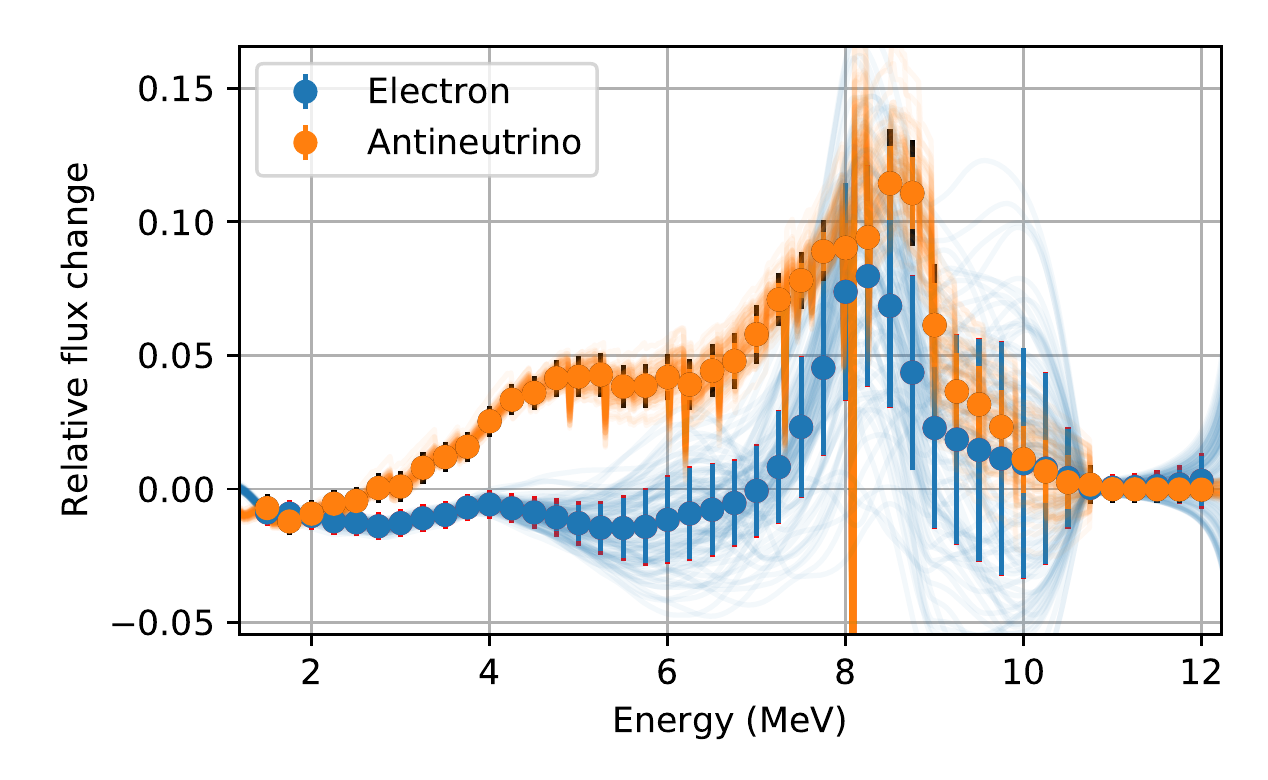}
    \vspace{-20pt}
    
    \includegraphics[width=0.48\textwidth]{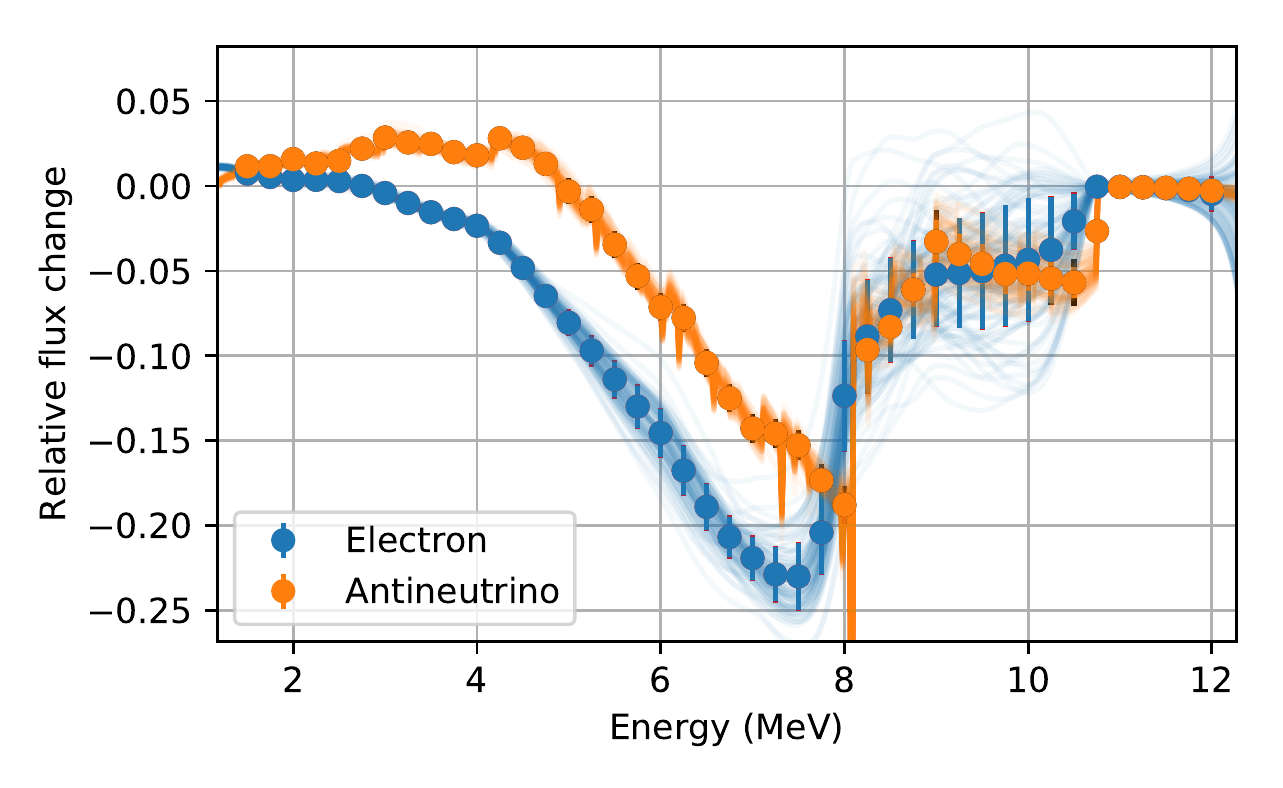}
    \caption{Spectral change for electron and antineutrino cumulative spectra in the pure summation approach using the methods discussed in the text for forbidden transitions. The energy axis shows the kinetic energy of the electron and antineutrino. Top panel: Comparison against the allowed transition with a weak magnetism term. Bottom panel: Comparison against treating all forbidden decays as unique. Uncertainties result from a Monte Carlo calculation of 100 samples, together with a theory uncertainty of 1\% from the uncertainty in the axial vector coupling constant, $g_A$, and pseudoscalar mesonic enhancement \cite{Hayen2018TBP}.}
    \label{fig:spectrum_change_Monte_Carlo}
\end{figure}

Figure \ref{fig:spectrum_change_Monte_Carlo} shows the spectral change and associated uncertainty for $^{235}$U in the so-called summation approach using 100 samples. We have not only made the comparison against the allowed approximation, but also against the Huber-Mueller prediction where all forbidden decays are treated as unique. We discuss both in turn.

As was observed already in the calculated results of Fig. \ref{fig:spectral_changes}, spectral changes to the electron cumulative spectrum are limited relative to the allowed approximation. The change in the antineutrino cumulative spectrum, on the other hand, shows significant deviations in the entire region of interest. Differences reach 5\% in the 5-6 MeV region, showing an increase of the predicted neutrino flux relative to the allowed approximation. The uncertainty shown is an uncorrelated combination of the theory uncertainty of 1\% due to the quenching of $g_A$ and mesonic corrections \cite{Hayen2018TBP, Kostensalo2018} and the Monte Carlo uncertainty. Compared to treating all forbidden decays as unique, on the other hand, significant deviations in both electron and antineutrino cumulative spectra are observed. Considering the large differences in shape (shown in Fig. \ref{fig:shape_factors_collection}) this is hardly surprising. This will be the subject of further research with relation to the reactor normalization anomaly.

The starting point of the usual anomaly and spectral shoulder analysis starts from a compatibility with the ILL data. In order to guarantee this agreement, we employ a mixed summation+conversion method as in Ref. \cite{Mueller2011}. Differences in calculated electron spectra from the summation component using our different approximations are then compensated by the conversion part of the procedure. Aside from $^{235}$U and $^{238}$U, however, summation predictions already exceed the experimental ILL data. For the $^{239,241}$Pu isotopes, then, the reference electron spectra are set to the summation calculation in the allowed approximation. As the implementation of forbidden transitions lowers the expected electron flux (see Fig. \ref{fig:spectral_changes}), this introduced deficit can be recovered analogously with the conversion procedure \cite{Hayen2018TBP}. The agreement with calculated and reference electron spectra is better than 1\% up to 7 MeV, after which the uncertainty in the calculated antineutrino spectra is linearly increased with the observed discrepancy in electron spectra.

By enforcing equivalence between electron spectra in our different approaches, the resultant antineutrino spectral changes can be directly compared to the shoulder observed experimentally. 

\begin{figure}[ht]
    \centering
    \includegraphics[width=0.5\textwidth]{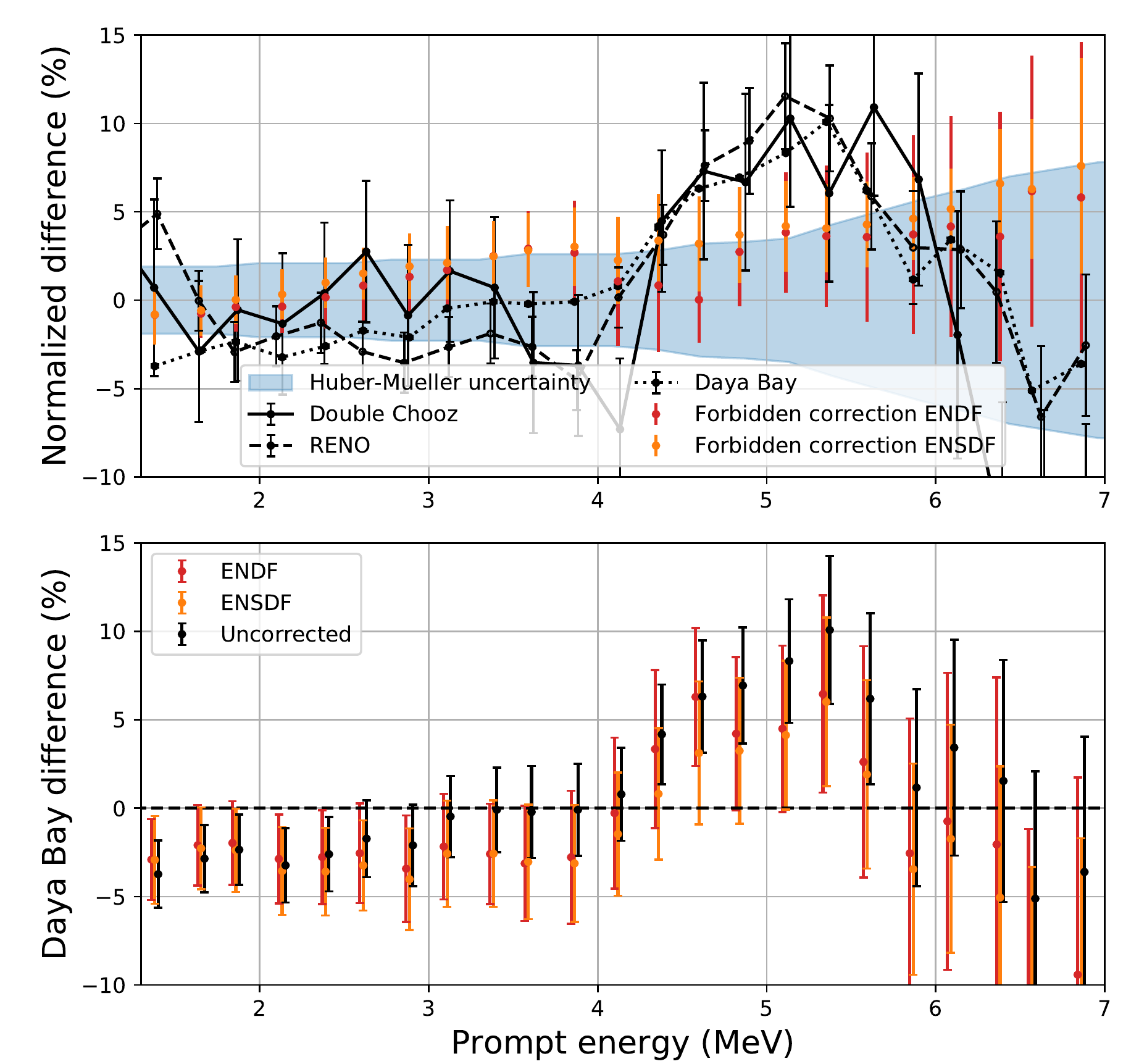}
    \caption{Top panel: Normalized spectral ratios for all three modern experiments relative to the Huber-Mueller predictions \cite{Mueller2011}, and the normalized forbidden spectrum correction described in this work using ENDF and ENSDF decay libraries. The prompt energy of the positron emerging from the inverse $\beta$ decay is related to the antineutrino energy via $E_\text{prompt} \approx E_\nu - 0.782$ MeV. The new results partially mitigate the original spectral shoulder and increase theoretical uncertainties from the treatment of first-forbidden decays. Bottom panel: Difference between Daya Bay spectral data and different theoretical models. Error bars are calculated using experimental, Huber-Mueller and forbidden uncertainties and are assumed uncorrelated. Here 'Uncorrected' is relative to the Huber-Mueller estimate shown in the top panel, and `ENDF' and `ENSDF' are the new results. For the latter two, almost all data points are consistent with zero within $1\sigma$.}
    \label{fig:reactors_ratio_comp}
\end{figure}

Figure \ref{fig:reactors_ratio_comp} shows the spectral ratio of Daya Bay \cite{An2016}, RENO \cite{Seo2018} and Double Chooz \cite{Abe2014} relative to the Huber-Mueller prediction with the uncertainty of the latter. Additionally, we show the correction from forbidden transitions as described above using a normalized spectrum between 2 and 8 MeV using the Daya Bay reactor composition \cite{An2012}, as is done for the the experimental results. Further, we show the discrepancy of the Daya Bay spectral data with respect to our new calculations. Due to the normalization requirement, the overestimate below 4 MeV is directly coupled to the underestimate in the bump region. The partial mitigation of the spectral shoulder and increased uncertainties arising from the treatment of first-forbidden transitions cause a significant reduction in the statistical significance. Based on the new results, the original spectral shoulder is now compatible with theoretical estimates as nearly all points agree within $1\sigma$.




In summary, we have for the first time performed microscopic calculations of the dominant forbidden transitions in the electron and antineutrino reactor spectra above 4 MeV. Through the use of a complete theoretical formalism, Coulomb corrections were taken into account at the appropriate level and shape factors strongly deviating from the usual allowed approximation were found. In combination with fission yield information, large changes were observed in the antineutrino spectrum. It was shown that, despite being limited in number, forbidden transitions are the dominant component of the electron flux between 2 and 7 MeV. Based on the uniform behaviour in the calculated shape factors, a parametrization of non-unique first-forbidden transitions was attempted. Using Monte Carlo methods, a spectral correction was obtained for all first-forbidden and higher uniquely forbidden transitions with an associated uncertainty. When compared to spectral discrepancies reported by all modern reactor neutrino experiments, the correction was shown to be of similar shape and magnitude. Taking these results at face value, a large portion of the reactor shoulder is mitigated. Due to increased theoretical uncertainties arising from an improved treatment of first-forbidden transitions, remaining spectral differences are statistically insignificant. Based on these results, it is clear that forbidden decays are not only non-negligible, but form an essential ingredient in the understanding of reactor antineutrino spectra and merit additional research.

\bibliography{library}

\end{document}